\documentclass[a4paper]{article}

\usepackage{INTERSPEECH2021}
\usepackage[dvipsnames]{xcolor}
\usepackage{tikz}
\usepackage{pgfplots}
\pgfplotsset{compat=1.16}
\usepackage{pgf-pie}
\usepackage{todonotes}
\usepackage{url}
\usepackage[T1]{fontenc}
\usepackage[utf8]{inputenc}
\usepackage{authblk}

\usepackage[all]{nowidow}

\definecolor{pie_undecided}{RGB}{222,235,247}
\definecolor{pie_incorrect}{RGB}{158,202,225}
\definecolor{pie_correct}{RGB}{49,130,189}

\definecolor{quote_green}{RGB}{0,100,0}

\definecolor{fluxcolor}{rgb}{0.9,0.6,0.1}

\definecolor{juliacolor}{rgb}{0.3,0.95,0.1}

\definecolor{nadjacolor}{rgb}{0.1,0.95,0.95}

\title{PoeticTTS - Controllable Poetry Reading for Literary Studies}

\name{Julia Koch$^1$, Florian Lux$^1$, Nadja Schauffler$^1$, Toni Bernhart$^2$, Felix Dieterle$^3$, Jonas Kuhn$^1$, \\ Sandra Richter$^{2,3}$, Gabriel Viehhauser$^2$, Ngoc Thang Vu$^1$}
\address{$^1$Institute for Natural Language Processing (IMS), University of Stuttgart, Germany \\
    $^2$Institute of Literary Studies (ILS), University of Stuttgart, Germany \\
    $^3$German Literature Archive (DLA), Marbach, Germany}
\email{julia.koch@ims.uni-stuttgart.de}
\begin{document}

\maketitle
\begin{abstract}
  Speech synthesis for poetry is challenging due to specific intonation patterns inherent to poetic speech. In this work, we propose an approach to synthesise poems with almost human like naturalness in order to enable literary scholars to systematically examine hypotheses on the interplay between text, spoken realisation, and the listener’s perception of poems. To meet these special requirements for literary studies, we resynthesise poems by cloning prosodic values from a human reference recitation, and afterwards make use of fine-grained prosody control to manipulate the synthetic speech in a human-in-the-loop setting to alter the recitation w.r.t. specific phenomena. We find that finetuning our TTS model on poetry captures poetic intonation patterns to a large extent which is beneficial for prosody cloning and manipulation and verify the success of our approach both in an objective evaluation as well as in human studies.
\end{abstract}
\noindent\textbf{Index Terms}: controllable speech synthesis, style cloning, human-in-the-loop

%
%
%
%
%
%

\section{Introduction}
In this paper, we present an approach for controllable lyric poetry synthesis. Poetic speech can be challenging for text-to-speech systems given that readers typically approach this genre differently than when they are reading prose, such as texts from newspapers, or novels \cite{Blohm/etal:2022}. Lyric poetry alludes to oral literary traditions, especially in the Romantic period in which our context is set. Poems are organised in verse and follow a more regulated form with parallel structures and more rigid alternations of stressed and unstressed syllables. It has been found that poetic speech differs from prosaic speech in a number of intonational patterns for which \cite{Byers:1979} coined the term "poetic intonation", such as e.g. short intonation units, more pauses, intonation units of relatively equal length, and repetition of pitch patterns (\cite{Byers:1979} \cite{Barney:1999}).
Given these differences, a synthesis model exclusively trained on prose data may struggle to produce these genre-specific patterns in the intended way. 
Our synthesis model is developed within the project »textklang« \cite{schauffler-EtAl:2022:LREC}, which is an interdisciplinary collaboration combining literary studies, digital humanities, computational linguistics, laboratory phonology and speech technology. The object of investigation is lyric poetry from the Romantic era, and one of the aims of the project is to develop methodologies and tools to enable researchers to systematically investigate the interplay between text, prosodic realisation in recitation, and the listener's perception and interpretation. 
Given the limited number of data on the one hand and the wealth of influencing linguistic and contextual factors contributing to differences in recitation on the other hand, speech synthesis plays a major role in evaluating hypotheses on the interrelation between the text and sound dimension of poems in a systematic and controlled way.
The challenge is to generate test items for perception experiments in which the respective prosodic parameters on which the hypothesis is based, can be manipulated without compromising the naturalness of the recording. This then allows for controlled perception experiments evaluating assumptions about the effects particular prosodic aspects in the recitation have on the listener. A second important role of speech synthesis within the project is to give access to particular prosodic realisations without violating copyrights of the original recordings. 
Considering the specific requirements on our TTS system, we are facing several challenges: First, our model has to capture the diverse range of prosodic values of poetic reading in order to generate speech with adequate expressiveness; second, we have to be able to exactly reproduce reference recitations; and third, we need functionality for fine-grained prosody manipulation. 
Applying TTS to the reading of poems is still an underexplored research area.
An early attempt was made by \cite{delmonte-prati-2014-sparsar} who performed a text analysis of poems to inform their TTS about prosody. However, they did not develop a TTS system for poetry reading, but relied on customization options of an off-the-shelf TTS system. More recently, a variety of approaches 
aim at modeling the style of the synthesised speech by mapping prosody of a reference into a latent space, by means of a reference encoder \cite{skerry2018towards}, \cite{wang2018style}, Variational Autoencoder \cite{zhang_style_vae:19} or inverse autoregressive flow structure \cite{an21b_interspeech}. However, style transfer has been explicitly applied to poetry only in \cite{an21b_interspeech} who integrate Chinese poetry as an example style in their experiments. 
Although these architectures successfully mimic the style of a reference, controllability is only possible on a global level which is not precise enough for our use case. Hence, we separate poetic style from prosody transfer by first finetuning our TTS model on poetry data to learn a general sense of poetic reading style and clone the exact prosody of a reference in a second step. An attempt on prosody cloning has been made in \cite{klimkov2019fine}, however, limited to single speaker TTS. Similar to ours, \cite{mohan2021ctrl} as used in \cite{torresquintero21adept} extract explicit values for duration, F$_0$ 
and energy from a reference and concatenate these with the encoder output. \cite{mohan2021ctrl} further show that these explicit values can be manipulated by a human-in-the-loop. However, they use an autoregessive model, while ours builds on FastSpeech 2 \cite{ren2020fastspeech} which models these prosodic values more explicitly. Phone-wise prosody manipulation has also been proposed in \cite{lancucki2021fastpitch}, however, without prosody cloning. 
Our contributions in this paper are as follows: 1) We present PoeticTTS: A text-to-speech approach to reading poetry that allows scholars to individually manipulate prosodic features of a particular realisation of a poem while attaining an almost human-like naturalness by cloning prosodic parameters from a reference recording. These properties make it a valuable tool for literary studies. 2) We investigate the impact of finetuning on poetry by comparing it to a model which is almost exclusively trained on prose to analyse how well poetic reading style can be learned from data. 3) While similar techniques for human-in-the-loop prosody control have been proposed before, to our best knowledge we are the first who perform a systematic evaluation of this method and apply it to a real-world use case. 
We recommend readers to listen to the samples provided on our demo page, which also links to an interactive demo.\footnote{\url{https://poetictts.github.io/}} All of our code is available open source.\footnote{\url{https://github.com/DigitalPhonetics/IMS-Toucan}}

\section{Proposed method}\label{sec:method}
\begin{figure}[ht]
    \centering
    \includegraphics[width=\linewidth]{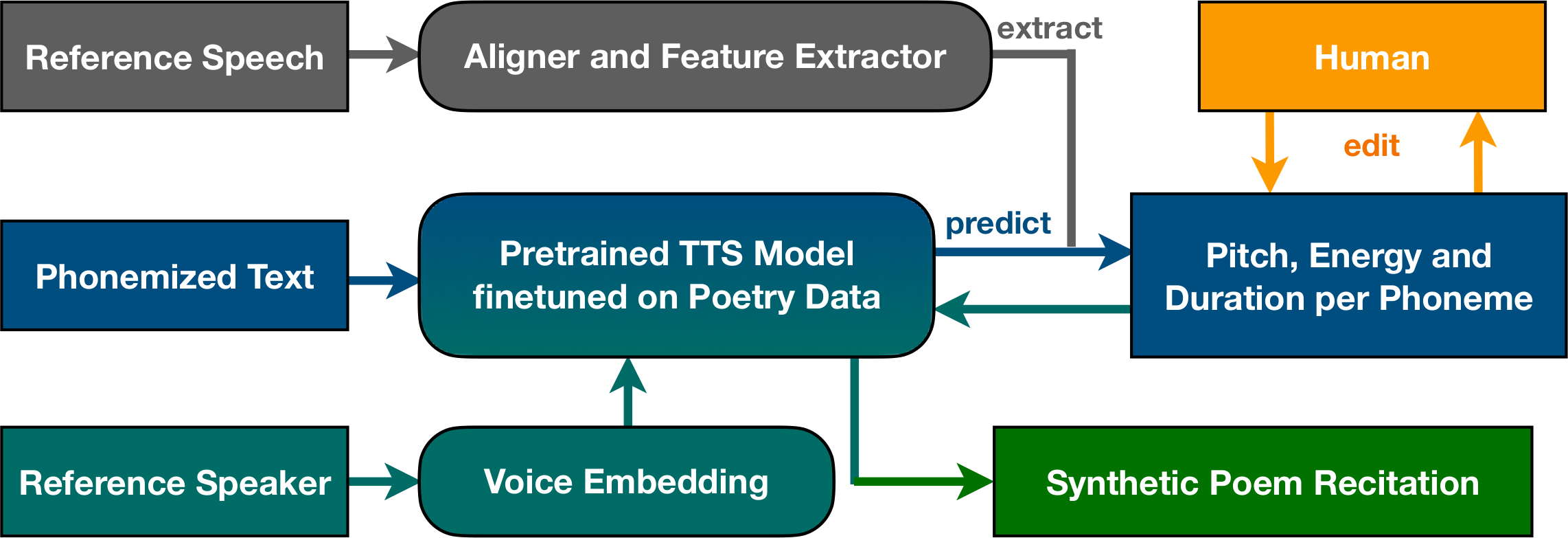}
    \caption{Overview of the proposed method. The reference extraction in gray overwrites the predictions, but is entirely optional. The actual speech production is marked in green.}
    \label{fig:overview}
\end{figure}

Our proposed method combines three components: First, we build a model that is capable to produce speech in a poetic reading style in a broad sense. We achieve this by leveraging poetic and non-poetic data during training. However, in order to provide a tool for our described use case in literary studies, it is necessary to have a prosodic replica of a specific poetic realisation. Hence, we clone the prosody from a reference audio to get an exact copy of a given recitation. Finally, we provide the opportunity to make human-in-the-loop adjustments to further manipulate the output. In the following, we describe each component in detail. Figure \ref{fig:overview} provides an overview of the modification pipeline.

\subsection{Model architecture and training procedure}
We use FastSpeech 2 \cite{ren2020fastspeech} together with a HiFi-GAN \cite{kong2020hifi} vocoder as our underlying model architecture as implemented in the IMS Toucan Toolkit \cite{lux2021toucan} with Conformer \cite{gulati2020conformer} blocks both in encoder and decoder. Further, this implementation uses FastPitch \cite{lancucki2021fastpitch} style phone-wise 
averaging of F$_0$ 
and energy values, which is the base for our controllability mechanism. To change the voice of the TTS to even unseen speakers, we condition the TTS on an ensemble of speaker embeddings \cite{xvect, ecapa} trained on Voxceleb 1 and 2 \cite{Chung18b, nagrani2017voxceleb, Nagrani19} using the Speechbrain toolkit \cite{speechbrain}. This approach is described in \cite{lux:2022:cloning} who show that prosody can be almost completely disentagled from other speaker characterisitcs. By this means, we can synthesise speech with the voice of an arbitrary reference speaker, which is not necessarily the same as the speaker of the reference recitation. This is required to factor out the influence of the speaker in our experiments. Following the LAML setup proposed in \cite{lux2022laml}, we pretrain our model on many hours of diverse speech in multiple languages and finetune a multispeaker model for German language on almost exclusively prosaic data. We use this as a basis to train our model in a second finetuning step on poetic data. Although we consider exclusively German data in this work, we find that our model tremendously benefits from multilingual pretraining such that it produces more expressive speech and is more robust to unseen prosodic variations in duration, F$_0$ 
and energy. 

\subsection{Prosody cloning}
Our method for prosody cloning follows the approach described in \cite{lux:2022:cloning}:
We extract ground-truth F$_0$
and energy values from a reference audio and in order to be transferable between different speakers, the values are normalized by dividing them by the average within the given reference, excluding zero values. For temporal alignment, we adopt \cite{lux:2022:cloning}'s reconstruction-based aligner which we trained together with our TTS model to predict the number of spectrogram frames for each phone.
To improve duration prediction, we finetune the aligner on the reference audio using Stochastic Gradient Descent before predicting durations at inference time. We then overwrite the predicted values for duration, F$_0$ 
and energy of the FastSpeech 2 encoder with the values extracted from the reference. 

\subsection{Human-in-the-loop manipulation}
In the same way as cloning duration, F$_0$ 
and energy from a reference, prosodic values can easily be redefined by a human expert. Manipulation of prosodic values can be done for each phone 
individually, since we average F$_0$ 
and energy values over all spectrogram frames that correspond to the phone. 
By first extracting the prosody of a reference audio and subsequent manual editing of the phone 
that add to a specific linguistic phenomenon, a human-in-the-loop can precisely control the realisation of this phenomenon while keeping everything else the same as in the original reference. While this is the intended usage of our system, we want to emphasise that prosody cloning and human-in-the-loop manipulation are independent from each other. In theory, it is possible to define the desired prosody completely manually without cloning from a reference first.

\section{Experimental setup}\label{sec:experiments}
We take our basic German FastSpeech 2 model before finetuning as baseline for our experiments to which we refer to as Prose model. We compare this to the model after finetuning on poetry, which we henceforth call Poetry model. To capture all aspects of our method in our experiments, we consider four model configurations: To examine the effects of finetuning alone, we compare both Prose and Poetry model without cloning prosody from a reference; and in order to evaluate the effectiveness of prosody cloning and manipulation, we compare both models with cloned prosody from a human reference.

\subsection{Data used}
We use a subset of the »textklang« corpus \cite{schauffler-EtAl:2022:LREC} as poetic data for evaluation and training of some of the models. In order to keep our data comparable throughout our experiments we focus on recitations of poems by  Friedrich Hölderlin. 
To train the aligner, vocoder and basic FastSpeech 2 system, we aimed for the greatest possible variance in training data, in order for the conditioning on the prosodic parameters to be most effective. Hence, we used a multilingual model configuration and trained on a total of 12 languages including the Blizzard Challenge 2011 dataset \cite{king2011blizzard},  LJSpeech  \cite{ljspeech17}, LibriTTS \cite{zen2019libritts}, HiFi-TTS \cite{bakhturina2021hi} and VCTK \cite{veaux2017superseded} for English, the HUI-Audio-Corpus-German \cite{puchtler2021huiaudiocorpusgerman} and the Thorsten corpus \cite{muller_thorsten_2021_5525342} for German, the Blizzard Challenge 2021 dataset \cite{ling2021blizzard} and the CSS10 dataset \cite{css10} for Spanish, as well as the CSS10 subsets for Greek, Finnish, French, Russian, Hungarian and Dutch, and further the Dutch, French, Polish, Portuguese and Italian subsets of the Multilingual LibriSpeech \cite{pratap2020mls}. To keep the computational cost manageable, we only use a maximum of 20,000 randomly chosen samples per corpus. This leaves us with a total amount of 400 hours of training data. The aligner is then finetuned for 10 steps on each individual sample before extracting durations from it. The multilingual FastSpeech 2 is finetuned for 500,000 additional steps on the German subset of the training data. From this, the Poetry model is further finetuned for 5,000 steps on 20 poems read by a single speaker. As this is very little data, we enhance our training data by automatically segmenting each poem into its stanzas as well as into single lines. This procedure also has the advantage, that our model sees long and short segments during finetuning to keep it flexible for variable length inputs at inference time. The vocoder does not require language or speaker specific finetuning at all.

\subsection{Objective evaluation metrics}


For an objective evaluation of our model, we compare synthesised speech generated by each of the four configurations to human references by calculating Log Mel Spectral Distortion (MSD) \cite{lux:2022:cloning}
with Dynamic Time Warping \cite{berndt1994using} and F$_0$ Frame Error (FFE) \cite{nakatani2008method} following \cite{skerry2018towards}. The MSD measures the euclidean distance between individual spectrogram frames in two sequences while finding the minimal average distance through DTW. So the lower the MSD score, the closer the two sequences match, while the two sequences do not need to have the exact same length. The FFE is calculated as the percentage of frames in which one or more pitch errors occur. Pitch errors are defined as 1) a deviation in F$_0$ 
value by more than 20\% and 2) an incorrect voicing decision. We use 95 stanzas from 17 unseen Hölderlin poems for evaluation.

\subsection{Poetic reading style}
 For a deeper analysis of the effects of finetuning, we test the ability of Poetry and Prose models without prosody cloning to read poetry with genre-appropriate intonation in a user study. We selected 8 excerpts of poems and 8 sentences from novels which cannot easily be identified as being prose or poetry from the text alone (e.g. no rhymes in the poem samples), and synthesised each set with both models.
 Participants were instructed to rate each sample as to whether it sounds more like poetry or more like prose. For comparability between the genres, all examples are taken from literature from the 19th century. 
 The synthesised samples are split into two groups such that each participant listens to each sentence only once, either read by the Prose or Poetry model. Hence, participants saw four samples for each combination of text genre and model, i.e. 16 
 in total.
\subsection{Human-in-the-loop}
We evaluate the usability of our human-in-the-loop setup in a second experiment. 
We use the Poetry model with cloned prosody to show and test this approach in an example research question on the realisation of enjambment. An enjambment occurs when the end of a 
verse disrupts a syntactic unit so that the line break suggests a prosodic boundary even though syntactically the sentence or clause continues, as shown below: \footnote{\textit{Let me at last, Father! with open eyes // face you! Have you not first [...]} (From: Friedrich Hölderlin, \textit{Der Zeitgeist})}

    \begin{quote}
      \guillemotright Laß endlich, Vater! offenen Aug's mich \textcolor{quote_green}{\textbf{dir
        Begegnen!}} hast denn du nicht zuerst [...] \guillemotleft
    \end{quote}

A speaker can deal with this conflict between discontinuation as suggested by the line break on the one hand and continuation as suggested by syntax on the other hand, in different ways, either emphasising the enjambment by realising a prosodic boundary at the end of the line, or conforming to syntactic continuity by reading over the line without employing phrasing cues. Some speakers may follow both ways by using cues typically found for marking prosodic boundaries, such as lengthening of the phrase-final segments, while at the same time signalling continuation by using cues typically found within phrases, e.g. F$_0$ downdrift and the absence of pauses (cf. \cite{Schauffler/etal:2022}). We can now control how an enjambment is realised in a given recitation by manipulating these cues, while leaving the surrounding context unchanged. 
As there are no gold-standard values to calculate objective metrics for this task, we evaluate the performance of our system in a second user study in an AXB setting. The purpose of this experiment is to show that our prosody cloning and human-in-the-loop approach is capable to make changes at the exact position specified and that our model realises the changes accurately. As it is difficult to define appropriate values for different realisations of the enjambments without solid expert knowledge, we select passages from poems for which we have two reference realisations by different speakers where one speaker emphasises the line break and the other one realises the enjambment with syntactic phrasing. We clone both realisations using the same voice embedding to generate reference audios A and B. We then generate a test sample X where we take the prosodic parameters of the enjambment from A as target values, while taking the surrounding context from reference B. In particular, we exchange the passage from the last word in the first line until the first pitch accent in the second line, as marked in the example above. 
To reduce complexity, we shorten the lines to complete syntactic units, i.e. in this example, we cut the passage after "Begegnen" which syntactically ends the sentence that started in the first line. We ask users to listen to both references as well as the test sample X and rate to which reference the realisation of the enjambment in X is more similar. We also show the text of the passage where we color-code the exchanged sequence and instruct them to pay particular attention to this area.
The study contains 14 items, each consisting of a test sample, a pair of two references, and the corresponding text.

\section{Results and discussion}
In the following, the MSD and FFE scores are presented first, followed by the human evaluation results of both experiments.

\subsection{Quantitative results}
Results for MSD and FFE are reported in Table \ref{tab:objective_results}. As expected, scores for both metrics are high for the uncloned configurations, i.e. where we synthesise the utterances with the values predicted by the model instead of extracting them from the reference. Nevertheless, we observe a reduction both in MSD and FFE of the Poetry model over the Prose model. We read from this that the Poetry model to some extent learned to imitate the prosody of the speaker reading poems, which leads to higher overlap between the spectrogram of the synthesised speech and the human reference, despite not knowing the actual prosodic values. Comparing the models in conjunction with prosody cloning, we still observe an advantage of the Poetry model over the Prose model, albeit to a lesser extent. This confirms that finetuning is a relevant step in our proposed method. 
Regarding the effectiveness of prosody cloning, we see substantial improvements of the model configurations using cloning over uncloned with a reduction of MSD by almost half in absolute numbers and comparable improvements for FFE. This shows that cloned speech is considerably more similar to the reference than without cloning, from which we conclude that our TTS system adopts the prosodic values from the reference adequately. This does not only show the success of our prosody cloning approach: We can also state that our approach is suitable for human-in-the-loop manipulation, since the mechanism for overwriting predicted values is the same as in prosody cloning.

\begin{table}[ht]
\caption{\label{tab:objective_results}
Mel Spectral Distortion and F$_0$ Frame Error for comparing the prosody of different TTS configurations to human-read poetry on 95 samples. The voice of the human speaker was cloned in all cases, since we are dealing with absolute F$_0$ 
values, which are speaker dependent.}
\centering
\begin{tabular}{ l | c | c } 
\noalign{\hrule height 1pt}
\textbf{Model Configuration}                   & \textbf{MSD} & \textbf{FFE} \\
\noalign{\hrule height 0.5pt}
Prose Model - uncloned \hspace{.5cm}      & 37.81           & 50.4\%   \\
Poetry Model - uncloned                   & 31.97           & 50.2\%   \\
Prose Model - cloned                          & 18.41           & 31.9\%   \\
Poetry Model - cloned                         & \textbf{17.53}           & \textbf{29.4\%}   \\
\noalign{\hrule height 1pt}
\end{tabular}
\end{table}  

\begin{figure}[ht]
    \centering
    \includegraphics[width=\linewidth]{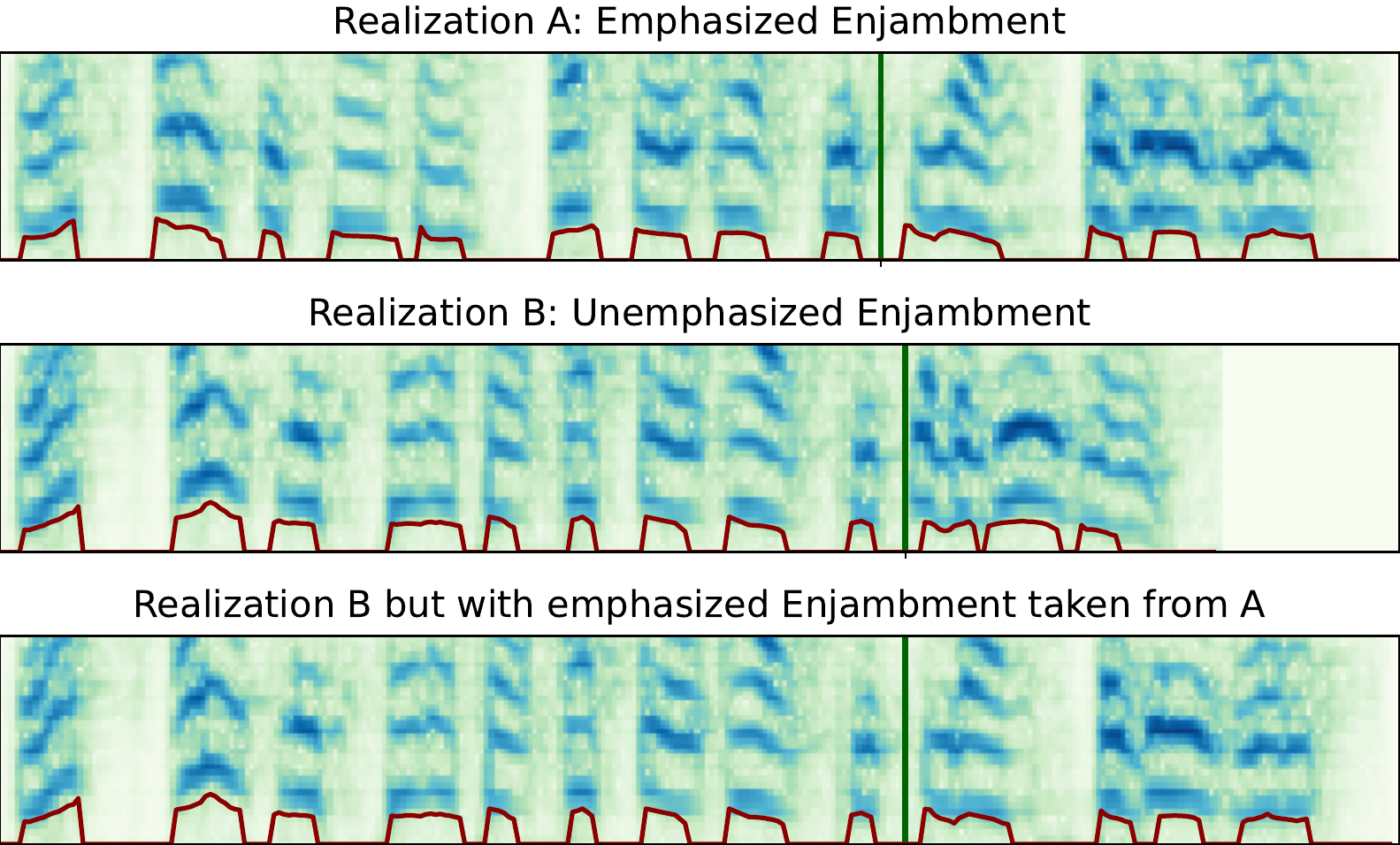}
    \caption{Exemplary comparison of two synthetic realizations of a poem and a third instance, in which the realization of the enjambment is taken from one reference, but everything else is taken from the other. For better visibility, the spectrogram only shows frequencies up to 500Hz. The pitch curve according to the Dio pitch tracker is displayed in red. The enjambment occurs at the end of the sample in this case and its beginning is marked with the green bar in each sample. The corresponding text is "Lass endlich Vater offenen Aug's mich dir // Begegnen!"}
    \label{fig:spectrogram}
\end{figure}

Figure \ref{fig:spectrogram} visualises the resulting speech after manipulating the realisation of the enjambment together with both references. With realisation B being the base for the manipulated sample, the plot shows that the pitch curve in the sample is similar to that in reference B until the position where the manipulation is done. This indicates, that human-in-the-loop manipulation at a specific position has no undesired side-effects on the areas that should not be affected. Regarding the realisation of the manipulated values, i.e. the passage to the right of the green bar in this example, the pitch curve closely resembles that of the target realisation A, showing that the manipulation was realised in the desired way by our system.

\subsection{Human evaluation}
\begin{figure}[ht]
    \centering
    \begin{tikzpicture}
 
    \begin{axis} [
        ybar = .08cm,
        height = 5cm,
        width = \linewidth,
        bar width = 32pt,
        ymin = 0,
        ymax = 85,
        symbolic x coords={Poetry Text,Prose Text},
        xtick distance = 1.0,
        ylabel = \small{Genre Perceived as Poetry in \%},
        nodes near coords,
        every node near coord/.append style={font=\tiny},
        enlarge x limits = 0.5,
        legend style={at={(0.65,1.0)},
        anchor=north,legend columns=-1},
        legend image code/.code={
        \draw [#1] (0cm,0cm) circle (0.1cm); },
    ]
 
    \addplot[fill=pie_correct,draw=white] coordinates {(Poetry Text,75.56) (Prose Text,63.33)};
    \addplot[fill=pie_incorrect,draw=white] coordinates {(Poetry Text,24.44) (Prose Text,16.67)};
 
    \legend {Poetry Model, Prose Model};
 
    \end{axis}
 
    \end{tikzpicture}
    \caption{Visualization of 720 ratings from 45 human raters on whether synthesised speech is read in a poetic reading style.}
    \label{fig:poem_or_not}
    \vspace*{-5mm}
\end{figure}
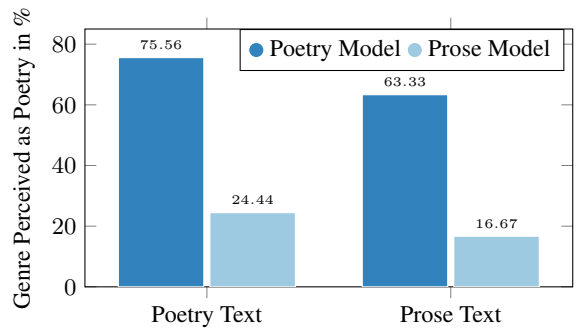

\paragraph*{Poetic reading style}
Figure \ref{fig:poem_or_not} shows the results of our human study on whether speech generated by the Poetry or Prose model realises poetic intonation appropriately. From 45 participants who took part in our study, we received a total of 180 answers for each of the four possible combinations of text genre and model. Summarising over all answers, speech generated by the Poetry model was perceived as poetic reading more often than for the Prose model by a large margin. The magnitude of this difference shows, that the choice of the model has a huge impact on the perception of speech, while the actual genre of a text plays only a subordinate role, supporting our hypothesis that finetuning is crucial to adapt to a poetic reading style.

\paragraph*{Human-in-the-loop}
We received a total of 462 answers from 33 participants for our second study. In 76.62\% of all answers participants were able to correctly identify the reference from which we took the prosodic parameters for the realisation of the enjambment. In the remaining cases, participants picked the wrong reference as answer (16.23\%) or could not decide for either of the references (7.14\%). Considering the difficulty of the task to identify subtle local differences 
while ignoring differences and similarities elsewhere, we see the vast majority of correct answers as evidence that our system realises the manipulated values precisely.

\section{Conclusions}
We presented PoeticTTS, an approach which enables TTS to read lyric poetry in a poetry-specific reading style. We achieve this by generating an exact copy of a given recitation by means of prosody cloning. Together with the possibility for human-in-the-loop manipulations, our approach can be a valuable tool in the context of literary studies. We have provided proof of concept for the combination of prosody cloning and human-in-the-loop manipulation by means of objective metrics as well as in human studies, and, additionally, we showed that TTS performance highly benefits from finetuning on genre-specific data. In future work, we want to include more diverse poetry from different authors and speakers into our approach and simplify the procedure for human-in-the-loop prosody manipulation by means of an intuitive user interface.

\section{Acknowledgements}
This research is supported by funding from the German Ministry for Education and Research (BMBF) for the »textklang« project.

\bibliographystyle{IEEEtran}

\bibliography{mybib}

\end{document}